# Signal attenuation and phase evolution evaluation under the influence of nonlinear gradient field

Chenghao Xu[a], Guoxing Lin[b]*

a. Mass Academy, Worcester, MA 01605, USA

b. Carlson School of Chemistry and Biochemistry, Clark University, Worcester, MA 01610, USA

**Abstract**

Accurately analyzing NMR and MRI diffusion experimental data relies on the theoretical expression used for signal attenuation or phase evolution. In a complex system, the encountered magnetic field is often inhomogeneous, which may be represented by a linear combination of $z^n$ gradient fields, where $n$ is the order. The current reported results on parabolic and cubic fields are inadequate to fully understand the effect of inhomogeneous fields. Additionally, the higher the order of the nonlinear gradient field, the more sensitive the phase variances are to differences in diffusion coefficients and delay times. Hence, studying higher-order fields is important for a better understanding of complex systems and designing improved experiments. The recently proposed phase diffusion method offers a general way to analyze the spin phase evolution of higher-order gradient fields using a recursive approach. This method is used and demonstrated in detail in this paper to determine the phase evolution in a quadric field ($n$ = 4), which acts as a key link in understanding higher-order fields. Three different types of phase evolution in the quadric gradient field are obtained. Moreover, a general signal attenuation expression $\frac{1}{\sqrt[2n]{1+\varepsilon\langle\emptyset_{Diff}^2\rangle}}$ is proposed to describe the signal attenuation for spin diffusion from the origin of the nonlinear gradient field, where $\langle\emptyset_{Diff}^2\rangle$ is the mean square diffusing phase, and $\varepsilon$ is a constant depending on $n$ and the pulse sequence. This approximation is based on the short gradient pulse (SGP) approximation but is extended to include the finite gradient pulse width (FGPW) effect by using the mean square phase. Compared to other forms of signal attenuation, such as Gaussian and Lorentzian, this method covers a broader range of attenuation, from small to relatively large. Additionally, this attenuation is easier to understand than the Mittag-Leffler function-based attenuation. The results, particularly the phase and signal attenuation expressions obtained in this paper, potentially advance PFG diffusion research in nonlinear gradient fields in NMR and MRI.

1. **Introduction**

The pulsed-field gradient (PFG) technique is widely used in nuclear magnetic resonance (NMR) and magnetic resonance imaging (MRI) [1,2,3,4,5,6]. In a magnetic field, the nuclear spin moment precession frequency $\omega(r)$ is proportional to the magnetic field strength $B(r)$ at the position $r$ where the spin is located, $\omega(r) = -\gamma B(r)$, where $\gamma$ is the gyromagnetic ratio. When a field gradient pulse is applied during experiments, the magnetic field becomes position-dependent; consequently, the spins' precession frequencies and their corresponding accumulated phase $\emptyset(r) = \int_0^{t_{tot}} \omega(r)dt$, are also position-dependent. This position-dependent accumulated phase can be employed to encode spatial information in MRI experiments. Additionally, different orders of spin quantum coherences evolve at different frequencies in a magnetic field. Using appropriate dephasing and rephasing gradient pulses, it is possible to selectively refocus on the desired coherence pathway, which makes the PFG an essential tool for selecting the coherence transfer pathway or suppressing undesired signals in modern NMR experiments.

*Email: glin@clarku.edu

Furthermore, for a diffusing spin, its precession frequency $\omega(r(t))$ varies along the random diffusion path, depending not only on position but also on time. This variation prevents the accumulated phases of diffusing spins from being refocused, even when dephasing and refocusing gradient pulses are applied. The spreading of the non-refocused accumulated phases results in signal attenuation or even a net phase shift of the observed magnetization in a gradient field in NMR and MRI experiments [5,6,7,8]. Analyzing changes in signal attenuation or phase shift can help determine diffusion parameters, such as the diffusion coefficient, which has critical applications in clinical diffusion MRI [9], as well as diffusion NMR (also known as NMR diffusion measurements, pulsed gradient spin echo (PGSE) NMR, or diffusion-ordered spectroscopy (DOSY) [10]).

The study of spin dynamics under linear or nonlinear magnetic field gradients remains a significant area of both theoretical and experimental research in NMR and MRI [5,6,8]. Although the linear magnetic field has been extensively examined and is widely utilized, nonlinear gradient fields occur in many situations [11,12,13,14,15]. The internal magnetic fields within actual samples, such as porous materials, are often inhomogeneous. These inhomogeneities can result from imperfections in the external magnetic field, eddy currents, as well as the sample's susceptibility and shape. Such magnetic field inhomogeneity can adversely impact experimental results, causing artifacts in diffusion imaging, reducing spatial resolution in MRI [16,17], and complicating the interpretation of NMR relaxation and diffusion data [18]. Often, these inhomogeneous fields are nonlinear. Accurate modeling of spin phase evolution is crucial for correctly interpreting signal attenuation and phase behavior in complex systems where the magnetic field varies nonlinearly with position. In these environments, traditional linear gradient models are insufficient, and higher-order effects must be considered.

Theoretical research on nonlinear gradients can not only offer a better understanding of complex systems in experiments but also assist in developing new experimental designs that utilize the advantages of the nonlinear field. Nonlinear gradient fields offer specific benefits over linear fields [19,20]. First, the net phase shift caused by the nonlinear field could be used to measure the diffusion coefficient directly. Second, for an $n$ order gradient field, the accumulated phase variance is approximately $\langle \phi^2 \rangle \propto D^n t^{n+2}$ [8]. The signal attenuation in higher-order gradient fields can potentially be more sensitive to changes in the diffusion coefficient and diffusion delay time than that in a linear gradient field. This sensitivity could be utilized to create experiments with enhanced MRI contrast based on variations in diffusion coefficients or to perform diffusion measurements with shorter delay times.

Unlike extensive theoretical research in the linear gradient field, studies on the nonlinear gradient field are limited. Nonlinear gradient fields have posed challenges for traditional theories [11-14,18,21,22], with limited and inconsistent results reported. However, advances have recently been made in this area. A recently proposed phase diffusion method not only resolves the discrepancies in traditional results but also shows that the diffusing spin system undergoes three types of phase evolutions under a nonlinear gradient field [23]. These evolutions are phase diffusion, float phase evolution, and shift evolution (the shift evolution depends on the initial position). The float phase depends on the contribution of the second-order derivative of the gradient field. Traditional theoretical approaches have not provided a clear understanding of the evolution of the float phase. Often, it is overlooked or misrepresented in reported theoretical NMR signal expressions, which hinders the accurate interpretation of experimental data. The float phase evolution can influence either phase shift, signal attenuation, or both, depending on the order of the nonlinear gradient field. In a parabolic field, it only affects phase shift, whereas in a cubic field, it impacts both phase shift and signal attenuation.

However, the phase diffusion method has only reported results for parabolic and cubic nonlinear gradient fields. In practical applications, besides these fields, there are other higher-order nonlinear fields ($n$ =4,5,6) commonly encountered in modern NMR spectroscopy, such as these compensational shimming fields. The z-direction gradient fields are typically denoted by $z^n, n$ =1-6 ($n$ =1 is the linear gradient field) [24]. The inhomogeneous field can often be decomposed into a linear combination of $z^n$ fields in NMR shimming [24]. The phase shift caused by the nonlinear field can be assessed by summing contributions from all terms in the linear combination. While the impact of signal attenuation from these terms is more complex, whose contributions appear in the superposition term of the coefficient of phase evolutions (such as $\sum_{n\geq 1}\left(K_n(t_{tot}) - K_n(t_j)\right)\left(z^n(t_j)\right)'$ in the diffusing phase evolution, where $K_n$ are wavenumbers) based on the phase diffusion method in Ref. [23]. As the gradient's complexity increases, so does the need for precise calculations of phase shift contributions from each term in the field expansion. The phase evolution of a higher-order gradient field can be determined based on the results of a lower-order gradient field, using a recursive method [8,23]. To fully understand the commonly encountered nonlinear field, it is insufficient to analyze only the parabolic and cubic fields; the fourth-order or even higher-order linear fields are necessary. The fourth-order gradient field in phase evolution acts as a key link in defining the behavior of even higher-order contributions. For simplicity, we only focus on a single order gradient field.

Additionally, a general type of signal attenuation function is proposed. Besides the previously reported Gaussian, Lorentzian, and MLF attenuations, a different type of signal attenuation function $\frac{1}{\sqrt[2n]{1+\varepsilon\langle\emptyset_{Diff}^2\rangle}}$ for diffusion from origin in an $n$-order nonlinear field is proposed based on the short gradient pulse (SGP) approximation results, where $\langle\emptyset_{Diff}^2\rangle$ is the mean square diffusing phase, and $\varepsilon$ is a constant depending on experiment sequence and gradient field order $n$. This attenuation is straightforwardly extended to include the finite gradient pulse width (FGPW) effect by using the mean square phase. The proposed attenuation expression shows good agreement with the simulation. It could have a broader applicable signal attenuation range than other types of attenuation. Furthermore, the general expression of the average phase shift for an $n$-order of the nonlinear field is given. The results provide additional insights into employing the phase diffusion method and analyzing spin self-diffusion under a nonlinear field, which could benefit our signal analysis and inform the design of new experiments.

**2. Theory**
2.1 General expressions of the accumulated phase
For simplicity, we only consider the one-dimensional gradient field along $z$ direction. The magnetic field $B(z)$ with a quartic gradient can be expressed as

$$B(z) = B_0 + g_n z^n, n = 4, \tag{1}$$

where $B_0$ is the static external field, and $g_4$ is the gradient constant. Based on expression (6a) in Ref. [23], the accumulated phase results from the quartic gradient field is,

$$\begin{aligned}\phi_4(t) &= \phi_D + \phi_{float} + \phi_{shift} \\ &= -\sum_{j=1}^{m}\left(K_4(t_{tot}) - K_4(t_j)\right)\left[f_4'\left(z(t_j)\right)\Delta z_j + \frac{1}{2}f_4''\left(z(t_j)\right)(\Delta z_j)^2\right] - \sum_{n\geq 1} K_4(t_{tot})f_n(z_0) \\ &= -\sum_{j=1}^{m}\left(K_4(t_{tot}) - K_4(t_j)\right)f_4'\left(z(t_j)\right)\Delta z_j \\ &\quad -\frac{1}{2}\sum_{j=1}^{m}\left(K_4(t_{tot}) - K_4(t_j)\right)f_4''\left(z(t_j)\right)(\Delta z_j)^2 - \sum_{n\geq 1} K_4(t_{tot})f_n(z_0),\end{aligned}$$

(2)

where $f_4(z(t)) = (z(t))^4$, $f_4'(z(t))$ and $f_4''(z(t))$ are the first and second order derivatives of $f_4(z(t))$ defined respectively by

$$f_4'(z(t)) = \frac{d}{dz}\left[(z(t))^4\right] = 4(z(t))^3 \tag{3a}$$

and

$$f_4''(z(t)) = \frac{d^2}{dz^2}\left[(z(t))^4\right] = 12(z(t))^2, \tag{3b}$$

and $\phi_D(t_{tot})$, $\phi_{float}(t_{tot})$, and $\phi_{shift,z_0}(t_{tot})$ correspond to the first, second, and third terms in the last line of Eq. (2), respectively. The Taylor expansion of $f_4(z(t))$ to the second order is employed in Eq. (2). The general expressions for these three types of phase evolutions are given separately in the following:

i. Diffusion phase, $\phi_D$

According to Ref. [23],

$$\langle \phi_D(t_{tot})^2 \rangle = 2 \int_0^{t_{tot}} D_\phi(t)\, dt, \tag{4a}$$

where

$$D_\phi(t) = [K_4(t_{tot}) - K_4(t)]^2 \langle [f_4'(z(t))]^2 \rangle D. \tag{4b}$$

Substituting Eq. (3a) into Eq. (4b), we obtain

$$D_\phi(t) = [K_n(t_{tot}) - K_n(t)]^2 \langle [f_4'(z(t))]^2 \rangle D$$

$$= 16D[K_n(t_{tot}) - K_n(t)]^2 \langle [z(t)]^6 \rangle, \tag{5a}$$

where [25]

$$\langle [z(t)]^6 \rangle = (Z_0)^6 + 30(Z_0)^4 Dt + 180(Z_0)^2(Dt)^2 + 120(Dt)^3. \tag{5b}$$

ii. Float phase $\phi_{float}$

From Ref. [23],

$$\phi_{float}(t_{tot}) = -\sum_{j=1}^{m}\left(K_4(t_{tot}) - K_4(t_j)\right)f_4''(z(t))D\Delta t_j, \tag{6}$$

and

$$v_{float}(t) = -[K_4(t_{tot}) - K_4(t)]f_4''(z(t))D, \tag{7}$$

where $D = \langle \frac{(\Delta z_j)^2}{2\Delta t_j} \rangle$.

The float phase $\phi_{float}(t_{tot})$ and $v_{float}(t)$ could also be understood from the asymmetric diffusion results [26,27]. From Eq. (2), the absolute step length from $f_4'(z(t_j))\Delta z_j$ is larger when $\Delta z_{j+1}$ is in the same sign as $z(t_j)$ than that of opposite signs, which appears to be an asymmetric random walk. The approximated jump length difference is $\zeta = -\frac{1}{2}[K(t_{tot}) - K(t_j)]f_4''(z(t))2\Delta z_j \Delta z_j$, where $\frac{1}{2}$ is used to reflect that on average, around 50% jumps are with either positive or negative $z(t_j)$. The velocity of asymmetric diffusion $v_{float,a}(t)$ [26,27]

$$v_{float,a}(t) \approx \frac{1}{2}\langle\frac{\zeta}{\Delta t_j}\rangle = -[K_4(t_{tot}) - K_4(t)]f_4''(z(t))D, \qquad (8)$$

which is the same as Eq. (7).

Based on Eqs. (2), (3b) and (7) [23],

$$\langle\emptyset_{float}(t_{tot})\rangle = \int_0^{t_{tot}} v_{float}(t)dt$$

$$= -\int_0^{t_{tot}} 12D[K_4(t_{tot}) - K_4(t)][z(t)]^2 dt, \qquad (9)$$

where

$$v_{float}(t) = -12D[K_4(t_{tot}) - K_4(t)][z(t)]^2 D. \qquad (10)$$

$\phi_{float}$ is a phase evolution, including the integral of $[z(t)]^2$, and it can be converted by setting [8,23] the following virtual parameters $\gamma'$ and $g'_2(t)$:

$$\gamma' = 1, \qquad (11a)$$

and

$$g'_2(t) = 12D[K_4(t_{tot}) - K_4(t)]. \qquad (11b)$$

Substituting $\gamma'$ and $g'_2(t)$ into Eq. (4a), we have

$$\emptyset_{float}(t_{tot}) = -\int_0^t \gamma' g'_2(t)[z(t)]^2 dt', \qquad (12)$$

which has the same format as the integral expression for calculating the accumulation phase under a parabolic field in Ref. [23]. The theoretical results of spin self-diffusion under a parabolic field with the condition $K_2'(t_{tot}) \neq 0$ can be employed to calculate $\emptyset_{float}(t_{tot})$ described by Eq. (12), which is referred to as a recursive calculating method [8], namely, the phase of higher order gradient can be calculated based on the result of the lower order gradient. The virtual wavenumber for the gradient $g'_2(t)$ is

$$K_2'(t_{tot}) = \int_0^t \gamma' g'_2(t) dt'. \qquad (13)$$

$K_2'^{(t_{tot})} \neq 0$ should hold, and based on the parabolic field results [23],

$$\phi_{float} = \phi'_D + \phi'_{float} + \phi'_{shift}, \qquad (14)$$

which includes three different phase evolutions: $\phi'_D$, $\phi'_{float}$, and $\phi'_{shift}$. Substituting Eq. (14) into Eq. (2), we get

$$\phi_4(t) = \phi_D + \phi_{float} + \phi_{shift} = \phi_D + (\phi'_D + \phi'_{float} + \phi'_{shift}) + \phi_{shift}. \qquad (15)$$

iii. Shift phase $\emptyset_{shift,Z_0}$

From Eq. (2),

$$\emptyset_{shift,Z_0}(t_{tot}) = -K_4(t_{tot})(Z_0)^4. \qquad (16)$$

The calculation of $\phi_4(t)$ is pretty tedious, so we calculate the $\phi_D(t_{tot}), \phi_{float}(t_{tot})$, and $\phi_{shift,z_0}(t_{tot})$ separately. Additionally, we will calculate two types of gradient pulse sequences, one with $K_4(t_{tot}) \neq 0$ (a steady gradient field, or simply referred to as one-pulse) and another with $K_4(t_{tot}) = 0$ (two gradient pulses, or simply referred to as two-pulses). The two-pulse case is calculated in section 2.2, while the one-pulse case is calculated in section 2.3. The wavenumbers used in the calculations are listed in Table 1.

**Table 1.** The wavenumbers for three different field gradient pulse sequences.

| one gradient pulse $K_4(t_{tot}) \neq 0$ | diffusion under $\frac{\pi}{2} - \delta$ RF pulse sequence with a steady gradient field, $t_{tot} = \delta$ $$K_4(t_{tot}) - K_4(t) = \gamma g_4 \delta - \gamma g_4 t.$$ |
|---|---|
| two gradient pulses $K_4(t_{tot}) = 0$ | diffusion under $\frac{\pi}{2} - \delta - \pi - \delta$ RF pulse sequence with a steady gradient field (PGSE or PGSTE, $\Delta = \delta$), $t_{tot} = 2\delta$ $$K_4(t) = \begin{cases} \gamma g_4 t, 0 \leq t \leq \delta, \\ \gamma g_4(2\delta - t), \delta < t \leq 2\delta. \end{cases}$$ |
|  | diffusion under pulsed gradient field (PGSE or PGSTE, $\Delta \geq \delta$), $t_{tot} = \Delta + \delta$ $$K_4(t) = \begin{cases} \gamma g_4 t, 0 \leq t \leq \delta, \\ \gamma g_4 \delta, \delta < t \leq \Delta, \\ \gamma g_4(\Delta + \delta - t), \Delta < t \leq \Delta + \delta. \end{cases}$$ |

2.2 Phase evolution under two gradient pulses, $K_4(t_{tot}) = 0$

From Table 1, for diffusion under two gradient pulsed gradient field pulses (PGSE or PGSTE, $\Delta \geq \delta$),

$$K_4(t_{tot}) = 0, \tag{17a}$$

and

$$K_4(t) = \begin{cases} \gamma g_4 t, 0 \leq t \leq \delta, \\ \gamma g_4 \delta, \delta < t \leq \Delta, \\ \gamma g_4(\Delta + \delta - t), \Delta < t \leq \Delta + \delta. \end{cases} \tag{17b}$$

  i.     Diffusing phase $\phi_D$

Eqs. (17) can be substituted into Eq. (5a) to give

$$D_\emptyset(t) = \begin{cases} 16D\gamma^2 g_4^2 t^2[(Z_0)^6 + 30(Z_0)^4 Dt + 180(Z_0)^2(Dt)^2 + 120(Dt)^3], 0 \leq t \leq \delta, \\ 16D\gamma^2 g_4^2 \delta^2[(Z_0)^6 + 30(Z_0)^4 Dt + 180(Z_0)^2(Dt)^2 + 120(Dt)^3], \delta < t \leq \Delta, \\ 16D\gamma^2 g_4^2(\Delta + \delta - t)^2[(Z_0)^6 + 30(Z_0)^4 Dt + 180(Z_0)^2(Dt)^2 + 120(Dt)^3], \Delta < t \leq \Delta + \delta. \end{cases} \tag{18}$$

Substituting $D_\emptyset(t)$ into Eq. (4a) yields

$$\langle \emptyset_D(t_{tot})^2 \rangle = \frac{32D\gamma^2 g_4^2 \delta^2}{3} A(D, Z_0, \Delta, \delta), \tag{19a}$$

where

$A(D, Z_0, \Delta, \delta) = 90D^3\Delta^4 + 120D^3\Delta^3\delta + 90D^3\Delta^2\delta^2 + 36D^3\Delta\delta^3 - 24D^3\delta^4 + 180D^2Z_0^2\Delta^3 + 180D^2Z_0^2\Delta^2\delta + 90D^2Z_0^2\Delta\delta^2 - 54D^2Z_0^2\delta^3 + 45DZ_0^4\Delta^2 + 30DZ_0^4\Delta\delta - 15DZ_0^4\delta^2 + 3Z_0^6\Delta - Z_0^6\delta.$  (19b)

For the case of no delay, $\Delta = \delta$, Eq. (19a) reduces to

$$\langle \emptyset_D(t_{tot})^2 \rangle = \frac{16D\gamma^2 g_n^2 \delta^3 [4(Z_0)^6 + 120D\delta(Z_0)^4 + 792D^2\delta^2(Z_0)^2 + 624D^3\delta^3]}{3}. \tag{20}$$

From Eqs. (19) and (20), it is evident that $\langle \emptyset^2 \rangle \propto D^4 t^6$. When $\delta \ll \Delta$, based on SGP approximation, Eq. (19) reduces to

$$\langle \emptyset_D(t_{tot})^2 \rangle_{SGP} = \frac{32 D\gamma^2 g_4^2 \delta^2}{3}(90 D^3 \Delta^4 + 180 D^2 Z_0^2 \Delta^3 + 45 D Z_0^4 \Delta^2 + 3 Z_0^6 \Delta). \tag{21}$$

ii. Float Phase $\emptyset_{float}$

From Eq. (11b) with $K_4(t_{tot}) = 0$, we have

$$g'_2(t) = -12 D K_4(t) = \begin{cases} -12 D\gamma g_4 t, & 0 \le t \le \delta, \\ -12 D\gamma g_4 \delta, & \delta < t \le \Delta, \\ -12 D\gamma g_4 (\Delta + \delta - t), & \Delta < t \le \Delta + \delta. \end{cases} \tag{22}$$

By substituting the values of $\gamma' = 1$ and $g'_n(t)$ into Eq. (13), it can be calculated that

$$K'_2(t) = \begin{cases} -6 D\gamma g_4 t^2, & 0 \le t \le \delta, \\ -12 D\gamma g_4 \delta t + 6 D\gamma g_4 \delta^2, & \delta < t \le \Delta, \\ 6 D\gamma g_4 (\Delta + \delta - t)^2 - 12 D\gamma g_4 \Delta \delta, & \Delta < t \le \Delta + \delta. \end{cases} \tag{23}$$

Now, $K'_2(t_{tot})$ can be calculated by substituting $t = \Delta + \delta$ into Eq. (23) to give

$$K'_2(t_{tot}) = -12 D\gamma g_4 \Delta \delta. \tag{24}$$

Based on Eqs. (23) and (24), it can be calculated that

$$K'_2(t_{tot}) - K'_2(t) = \begin{cases} -12 D\gamma g_4 \Delta \delta + 6 D\gamma g_4 t^2, & 0 \le t \le \delta, \\ -12 D\gamma g_4 \Delta \delta + 12 D\gamma g_4 \delta t - 6 D\gamma g_4 \delta^2, & \delta \le t \le \Delta, \\ -6 D\gamma g_4 (\Delta + \delta - t)^2, & \Delta \le t \le \Delta + \delta. \end{cases} \tag{25}$$

From Eq. (14), $\emptyset_{float}(t_{tot})$ includes three components: $\phi'_D$, $\emptyset'_{float}$, and $\emptyset'_{shift}$, which are calculated below:

a. Diffusion component of float Phase, $\emptyset'_D$

The diffusion coefficient $D'_\emptyset(t)$ for the diffusion phase of $\emptyset_{float}(t_{tot})$ can be written as [23]

$$D'_\emptyset(t) = [K'_2(t_{tot}) - K'_2(t)]^2 \langle \{[z(t)]^2\}'\rangle^2 D$$

$$= [K'_2(t_{tot}) - K'_2(t)]^2 \langle \{2[z(t)]^1\}^2 \rangle D$$

$$= 4[K'_2(t_{tot}) - K'_2(t)]^2 \langle [z(t)]^2 \rangle D, \tag{26}$$

where $\langle [z(t)]^2 \rangle = (Z_0)^2 + 2Dt$. Substituting Eq. (25) into Eq. (26), we obtain

$$D'_\emptyset(t) = \begin{cases} 4D[(-12 D\gamma g_4 \Delta \delta) + 6 D\gamma g_4 t^2]^2 [(Z_0)^2 + 2Dt], & 0 \le t \le \delta, \\ 4D[-12 D\gamma g_4 \Delta \delta + 12 D\gamma g_4 \delta t - 6 D\gamma g_4 \delta^2]^2 [(Z_0)^2 + 2Dt], & \delta < t \le \Delta, \\ 4D[6 D\gamma g_4 (\Delta + \delta - t)^2]^2 [(Z_0)^2 + 2Dt], & \Delta < t \le \Delta + \delta. \end{cases} \tag{27}$$

Like the derivation of the diffusion term $\phi_D$ in Eq. (4) and in Ref. [23] for the parabolic field, $D'_\emptyset(t)$ can be used to calculate the phase variance

$$\langle \emptyset'_D(t_{tot})^2 \rangle = 2\int_0^{t_{tot}} D'_\emptyset(t)\, dt = \frac{96 D^3 \gamma^2 g_4^2}{5}(D\delta^6 - 4D\Delta\delta^5 + 15D\Delta^2\delta^4 + 20D\Delta^3\delta^3 + 10D\Delta^4\delta^2 +$$

$$20\Delta^3\delta^2 Z_0^2 + 30\Delta^2\delta^3 Z_0^2 - 5\Delta\delta^4 Z_0^2 + \delta^5 Z_0^2). \tag{28}$$

For the case of no delay, $\Delta = \delta$, Eq. (28) reduces to

$$\langle \phi'_D(t_{tot})^2 \rangle = \frac{96 D^3 \gamma^2 \delta^5 g_4{}^2 [46(Z_0)^2 + 42 D\delta]}{5}. \tag{29}$$

When $\delta \ll \Delta$, based on SGP approximation, Eq. (28) reduces to

$$\langle \phi'_D(t_{tot})^2 \rangle_{SGP} = 192 D^3 \gamma^2 g_4^2 \delta^2 (D\Delta^4 + 2\Delta^3 Z_0^2). \tag{30}$$

b. $\phi'_{float}$

From the float phase of the parabolic field reported in Ref [23], we have

$$\langle \phi'_{float}(t_{tot}) \rangle = \int_0^{t_{tot}} v'_{float}(t) dt, \tag{31}$$

where

$$v'_{float}(t) = -[K'_2(t_{tot}) - K'_2(t)] 2(2-1)[z(t)]^{2-2} D$$
$$= -2D[K'_2(t_{tot}) - K'_2(t)]. \tag{32}$$

By substituting Eq. (25) into Eq. (32),

$$v'_{float}(t) = \begin{cases} -2D[-12 D\gamma g_4 \Delta\delta + 6 D\gamma g_4 t^2], 0 \le t \le \delta, \\ -2D[-12 D\gamma g_4 \Delta\delta + 12 D\gamma g_4 \delta t - 6 D\gamma g_4 \delta^2], \delta < t \le \Delta, \\ -2D[-6 D\gamma g_4 (\Delta + \delta - t)^2], \Delta < t \le \Delta + \delta. \end{cases} \tag{33}$$

By substituting Eq. (33) into Eq. (31), the integral yields

$$\langle \phi'_{float}(t_{tot}) \rangle = 12 D^2 g_4 \gamma \delta (\delta\Delta + \Delta^2). \tag{34}$$

When $\Delta = \delta$,

$$\langle \phi'_{float}(t_{tot}) \rangle = 24 D^2 g_4 \gamma \, \delta^3. \tag{35}$$

When $\delta \ll \Delta$, based on the SGP approximation, Eq. (34) reduces to

$$\langle \phi'_{float}(t_{tot}) \rangle_{SGP} = 12 D^2 g_4 \gamma \delta \Delta^2. \tag{36}$$

$\phi'_{float}(t_{tot})$ is a pure phase shift term, therefore [23],

$$\langle [\phi'_{float}(t_{tot})]^2 \rangle = [\langle \phi'_{float}(t_{tot}) \rangle]^2 = [12 D^2 g_4 \gamma \delta (\delta\Delta + \Delta^2)]^2. \tag{37}$$

When $\delta \ll \Delta$, based on SGP approximation, Eq. (36) reduces to

$$\langle [\phi'_{float}(t_{tot})]^2 \rangle_{SGP} = [12 D^2 g_4 \gamma \delta \Delta^2]^2. \tag{38}$$

c. $\phi'_{shift,Z_0}$

According to Ref. [23],

$$\langle \phi'_{shift,Z_0}(t_{tot}) \rangle = -K'_2(t_{tot})(Z_0)^2 = 12 D\gamma g_4 \Delta\delta (Z_0)^2, \tag{39}$$

where $K'_2(t_{tot})$ is defined in Eq. (24). Eq. (39) does not change under the SGP approximation.

iii. $\phi_{shift,Z_0}(t_{tot})$

According to Ref. [23],

$$\langle \phi_{shift,Z_0}(t_{tot}) \rangle = -K_4(t_{tot})(Z_0)^4 = 0, \tag{40}$$

because $K_4(t_{tot}) = 0$ in the two gradient pulse case.

2.3 Phase evolution under one gradient pulse, $K_4(t_{tot}) \neq 0$

From Table 1, for diffusion under $\frac{\pi}{2} - \delta$ radio frequency (RF) pulse sequence with a steady gradient field, $t_{tot} = \delta$,

$$K_4(t_{tot}) = \gamma g_4 \delta, \tag{41}$$

and

$$K_4(t_{tot}) - K_4(t) = \gamma g_4 \delta - \gamma g_4 t. \tag{42}$$

    i.      Diffusing phase $\phi_D$

Eq. (42) can be substituted into Eq. (5a) to give

$$D_\phi(t) = 16D[\gamma g_4 \delta - \gamma g_4 t]^2 \langle [z(t)]^6 \rangle, \tag{43}$$

which can be substituted into Eq. (4a) to give

$$\langle \phi_D(t_{tot})^2 \rangle = 2\int_0^{t_{tot}} 16D[\gamma g_4 \delta - \gamma g_4 t]^2 \langle [z(t)]^6 \rangle \, dt. \tag{44}$$

$\langle [z(t)]^6 \rangle$ is given by Eq. (5b), which can be substituted into Eq. (44) to give

$$\langle \phi_D(t_{tot})^2 \rangle = 32D \int_0^{t_{tot}} [\gamma g_4 \delta - \gamma g_4 t]^2 \langle (Z_0)^6 + 30(Z_0)^4 Dt + 180(Z_0)^2 (Dt)^2 + 120(Dt)^3 \rangle \, dt$$

$$= \frac{16D\gamma^2 g_4^2 \delta^3 [2(Z_0)^6 + 15D\delta(Z_0)^4 + 36D^2\delta^2(Z_0)^2 + 12D^3\delta^3]}{3}. \tag{45}$$

    ii.     Float phase

Based on Eqs. (11b) and (42),

$$g'_2(t) = 12D[K_4(t_{tot}) - K_4(t)] = 12D[\gamma g_4 \delta - \gamma g_4 t], \tag{46}$$

which can be substituted into Eq. (13) to give

$$K_2'(t) = 12D\gamma g_4 \delta t - 6D\gamma g_4 t^2, 0 \leq t \leq \delta. \tag{47}$$

From Eq. (47), at $t = t_{tot} = \delta$,

$$K_2'(t_{tot}) = 6D\gamma g_4 \delta^2. \tag{48}$$

$$K_2'(t_{tot}) - K_2'(t) = 6D\gamma g_4 \delta^2 - 12D\gamma g_4 \delta t + 6D\gamma g_4 t^2. \tag{49}$$

    a.    Diffusion component of float phase

$$D'_\phi(t) = 4[K_2'(t_{tot}) - K_2'(t)]^2 \langle [z(t)]^2 \rangle D$$

$$= 4(6D\gamma g_4 \delta^2 - 12D\gamma g_4 \delta t + 6D\gamma g_4 t^2)[(Z_0)^2 + 2Dt]D, \tag{50}$$

which can be substituted into Eq. (4a) to give

$$\langle \phi'_D(t_{tot})^2 \rangle = 2\int_0^{t_{tot}} D'_\phi(t) \, dt = \frac{96D^3 \gamma^2 \delta^5 g_4^2 (3z_0^2 + D\delta)}{5}. \tag{51}$$

    b.    $\phi'_{float}$

Similarly to Eq. (32), based on the parabolic field result, we have

$$v'_{float}(t) = -[K'_2(t_{tot}) - K'_2(t)]2(2-1)[z(t)]^{2-2}D$$
$$= -12[D\gamma g_4 \delta^2 - 2D\gamma g_4 \delta t + D\gamma g_4 t^2]D, \quad (52)$$

and

$$\langle \emptyset'_{float}(t_{tot})\rangle = \int_0^{t_{tot}} v'_{float}(t)dt = -4D^2\gamma g_4 \delta^3. \quad (53)$$

Because $\emptyset'_{float}(t_{tot})$ is a pure phase shift term, we have [23]

$$\langle[\emptyset'_{float}(t_{tot})]^2\rangle = [\langle\emptyset'_{float}(t_{tot})\rangle]^2 = 16D^4\gamma^2 g_4^2 \delta^6. \quad (54)$$

c. $\emptyset'_{shift}$

According to Ref. [23], and based on Eq. (48), we have

$$\langle\emptyset'_{shift,Z_0}(t_{tot})\rangle = -K'_2(t_{tot})(Z_0)^2 = -6D\gamma g_4 \delta^2 (Z_0)^2. \quad (55)$$

iii. Shift phase

Based on Eqs. (16) and (41)

$$\langle\emptyset_{shift,Z_0}(t_{tot})\rangle = -K_4(t_{tot})(Z_0)^4 = -\gamma g_4 \delta (Z_0)^4. \quad (56)$$

2.4 The signal phase and signal attenuation expressions

The NMR signal is the average magnetization for spins starting from $z_0$.

$$S(t_{tot}) = exp\{i\langle\emptyset\rangle\}|S(t_{tot})|, \quad (57)$$

where $\langle\emptyset\rangle$ is the average phase, and $|S(t_{tot})|$ is the amplitude of the signal. Affected by a nonlinear gradient field, the phase distribution of the spin system is non-Gaussian, and the signal attenuation deviates from the exponential attenuation based on a Gaussian phase distribution. In Ref. [23], the Lorentzian phase distribution and heavy-tailed distribution are proposed to approximately describe the phase distribution, which yield non-Gaussian signal attenuation, such as Lorentzian or MLF attenuation. Here, we propose an alternative signal attenuation expression based on the SGP approximation, and we will generalize it to the non-SGP approximation situation.

2.4.1 Phase distribution and signal attenuation based on SGP approximation

The SGP approximation assumes that the duration of the applied gradient pulse, $\delta$, is very short and negligible, but the gradient is sufficiently strong to create the necessary phase spreading. Under an ideal situation, for a finite $\gamma g_n \delta$, $\delta$ can be assumed to be infinitely small, and $g_n$ is infinitely large; the diffusion during the infinitely small $\delta$ can be neglected. The phase distribution for diffusion in a two-pulse experiment with $n$-order gradient field can be described as

$$P_{SGP}(\emptyset) = \begin{cases} \frac{1}{\sqrt{4\pi Dt}} exp\left(-\frac{(\frac{\emptyset}{\gamma g_n})^{\frac{2}{n}}}{4Dt}\right) \frac{1}{n\gamma g_n (\frac{|\emptyset|}{\gamma g_n})^{1-\frac{1}{n}}}, odd\ n, \emptyset \in R, \\ \frac{2}{\sqrt{4\pi Dt}} exp\left(-\frac{(\frac{\emptyset}{\gamma g_n})^{\frac{2}{n}}}{4Dt}\right) \frac{1}{n\gamma g_n (\frac{\emptyset}{\gamma g_n})^{1-\frac{1}{n}}}, even\ n, \emptyset \geq 0. \end{cases} \quad (58)$$

Except for $n = 1$, $P_{SGP}(\emptyset)$ described by Eq. (58) is obviously non-Gaussian. Ref. [23] proposed Lorentzian and heavy tail distributions to approximately describe the phase distribution, and it found that the MLF function offers a better fit to simulation data than Gaussian and Lorentzian attenuations when signal attenuation is not too small.

Based on the phase distribution Eq. (23) obtained from the SGP approximation. The signal attenuation can be evaluated based on

$$|S_{SGP}(t_{tot})| = \begin{cases} \int_{-\infty}^{\infty} \cos(\emptyset) P_{SGP}(\emptyset) \, d\emptyset, & odd\ n, \\ \sqrt{\left[\int_0^{\infty} \sin(\emptyset) P_{SGP}(\emptyset) \, d\emptyset\right]^2 + \left[\int_0^{\infty} \cos(\emptyset) P_{SGP}(\emptyset) \, d\emptyset\right]^2}, & even\ n. \end{cases} \quad (59)$$

When $n = 1$, Eq. (59) gives $|S_{SGP}(t_{tot})| = exp\left\{-\frac{(\gamma g \delta)^2 Dt}{2}\right\} = exp\left\{-\frac{\langle\emptyset\rangle^2}{2}\right\}$, which is the familiar result for SGP under a linear gradient field. While, for $n = 2$, we get

$$|S_{SGP}(t_{tot})| = \frac{1}{\sqrt[4]{1+16(\gamma g \delta Dt)^2}} = \frac{1}{\sqrt[4]{1+2\langle\emptyset_{Diff}^2\rangle}}, \quad (60)$$

where $\langle\emptyset_{Diff}\rangle^2$ corresponds to the sum of diffusing phases from both the diffusion phase and float phase; $\langle\emptyset_{Diff}\rangle^2 = \langle\emptyset_D\rangle^2 = 8(\gamma g \delta Dt)^2$ has been reported for diffusion in a parabolic field in Ref. [23]. In a parabolic field, the float phase has no diffusing phase components.

However, for $n \geq 3$, the integrals in Eq. (59) are complicated, and they currently do not have closed forms. Eq. (59) can be numerically evaluated. The following expression is found that could describe the signal attenuation of the numerical evaluation and simulation results:

$$|S(t_{tot})| \approx \frac{1}{\sqrt[2n]{1+\varepsilon\langle\emptyset_{Diff}^2\rangle}}, \quad (61)$$

where $\langle\emptyset_{Diff}^2\rangle$ is the sum of all diffusing phase contributions from diffusion and float phase evolutions, and $\varepsilon$ is constant. Although Eq. (61) is obtained based on the SGP approximation, it can be straightforwardly generalized to include the finite gradient pulse width (FGPW) effect, just using $\langle\emptyset_{Diff}^2\rangle$ calculated with the FGPW effect. This generalization is reasonable, considering the phase distributions in both SGP and FGPW cases are similar, because their corresponding phase evolution obeys the same type of equation, Eq. (2).

Fig. 1 shows the comparison between the predicted curve based on Eq. (61) and the numerical evaluation results, for $n = 3,4$. From Fig. 1, $\varepsilon$ equals 2 and 1/4 for $n = 3,4$, respectively. The value of $\varepsilon$ differs in the one-pulse and two-pulse experiments for $n = 4$ as shown in Section 3. There is good agreement between the theoretical prediction based on Eq. (61) and the numerical integration based on Eq. (59).

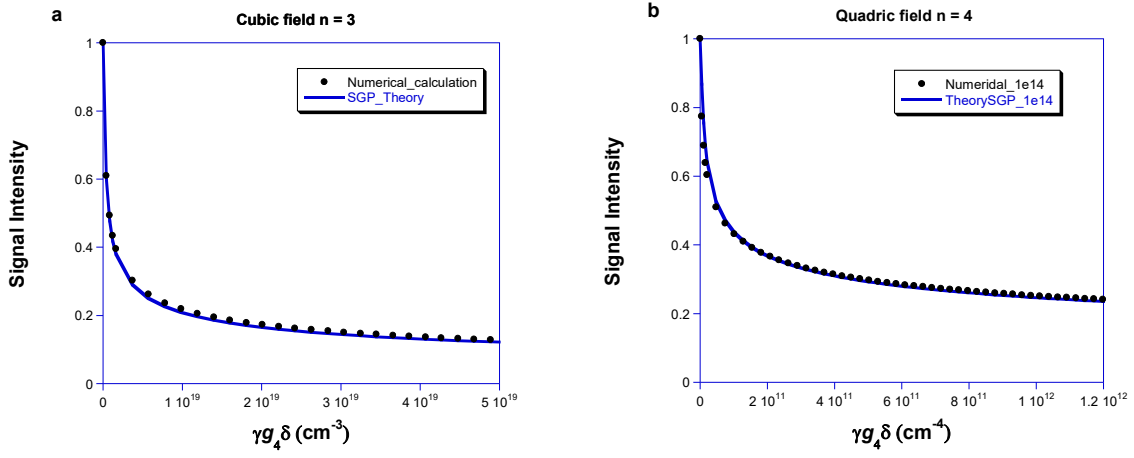

**Fig. 1** Comparison of theoretical NMR signal attenuations with numerical evaluations: (a), cubic field, (b), quadric field. The numerical evaluation is obtained based on Eq. (59). The theoretical predictions are based on Eq. (61), $\frac{1}{\sqrt[2n]{1+\varepsilon\langle\phi_{Diff}^2\rangle}}$, where $\varepsilon$ equals 2 and 1/4 are used for $n = 3,4$, respectively.

2.4.2 General phase and signal expression for a quadric field, $n = 4$

Ref. [20] proposed a general method to evaluate the mean phase shift for diffusion starting from the origin under the influence of a parabolic field. This method has been generalized in Ref. [23] to spin diffusion beginning from a random location. Here, this method is further generalized to calculate the phase of a random-order nonlinear gradient field. The average phase shift of spin diffusion under an n-order nonlinear gradient field can be calculated by

$$\langle\phi\rangle = \int_{-\infty}^{\infty} \phi\, P(\phi)d\phi = -\int_0^{t_{tot}} \gamma g_n(t)\left[\int_{-\infty}^{\infty}(z(t))^n \frac{1}{\sqrt{4\pi Dt}}exp\left(-\frac{(z(t)-z_0)^2}{4Dt}\right)\right]dt$$

$$= -\int_0^{t_{tot}} \gamma g_n(t)\langle(z(t))^n\rangle dt, \tag{62}$$

where $\langle(z(t))^n\rangle$ is the $n$th moment of the probability distribution. For $n = 4$, $(z(t))^4 = 12(Dt)^2 + 12Dtz_0^2 + z_0^4$

$$\langle\phi\rangle = -\int_0^{t_{tot}} \gamma g_4(t)\langle(z(t))^4\rangle dt \xrightarrow{K_4(t_{tot})=0,\,(z(t))^4=12(Dt)^2+12Dtz_0^2+z_0^4}$$
$$12\gamma g_4\delta D^2\Delta^2 + 12\gamma g_4\delta D^2\Delta\delta + 12\gamma g_4\delta z_0^2 D\Delta, \tag{63}$$

where $K_4(t_{tot}) = 0$ indicates that in the dephasing and refocusing pulses, the phases are opposite. Unlike the parabolic field, the expression of the phase shift for quadric field depends on $z_0^2$. When $Z_0 = 0$,

$$\langle\phi\rangle = 12\gamma g_4\delta D^2\Delta^2 + 12\gamma g_4\delta z_0^2 D\Delta. \tag{64}$$

The average phase from Eq. (64) equals the sum of $\phi'_{float}, \phi'_{shift,Z_0}$, and $\phi_{shift,Z_0}$.

$$\langle\phi\rangle = \phi'_{float}(t_{tot}) + \phi'_{shift,Z_0}(t_{tot}) + \phi_{shift,Z_0}(t_{tot}). \tag{65}$$

Under the quadric field, both $\phi_D$ and $\phi'_D$ are the diffusing phase, which affects the signal amplitude $|S(t_{tot})|$, while $\phi'_{float}(t_{tot})$, $\phi'_{shift,Z_0}(t_{tot})$, and $\phi_{shift,Z_0}(t_{tot})$ are phase shifts without distributions, which affect the total signal phase. Therefore, the NMR signal can be expressed as

$$S(t_{tot}) = exp\{i \langle \phi \rangle\}|S(t_{tot})| = exp\{i [\phi'_{float}(t_{tot}) + \phi'_{shift,Z_0}(t_{tot}) + \phi_{shift,Z_0}(t_{tot})]\}|S(t_{tot})|, \quad (66)$$

where

$$|S(t_{tot})| = exp\left\{-\frac{[\langle(\phi_D(t_{tot}))^2\rangle + \langle(\phi'_D(t_{tot}))^2\rangle]}{2}\right\}, \text{ Gaussian phase distribution}, \quad (67)$$

which is based on the assumption that $\phi_D$ and $\phi'_D$ follow Gaussian distributions. Eq. (67) works for small signal attenuation and diffusion starting far away from the origin of the gradient field (the phase distribution can be approximated as Gaussian for $z_0 \gg Dt$).

However, the correlation between the coefficients of the individual jump steps of the phase diffusion $\phi_{float}(t_{tot})$ and $\phi_D(t_{tot})$ should make the diffusion deviate from Gaussian diffusion. Based on the SGP approximation, and non-Gaussian type distributions: Lorentzian distribution and long-tailed phase distribution proposed in Ref. [23], the amplitude of the signal attenuation $|S(t_{tot})|$ in Eqs. (26a) and (26b) will be replaced as

$$|S(t_{tot})| = \begin{cases} exp(-\Upsilon(t_{tot})), \text{ Lorentzian phase distribution}, \\ E_\alpha(-\Upsilon(t_{tot})), \text{ long-tailed fractional phase distribution}, \\ \frac{1}{\sqrt[2n]{1+\varepsilon\langle\phi^2_{Diff}\rangle}}, \text{ based on SGP approximation}, \end{cases} \quad (68a)$$

where $E_\alpha(-\Upsilon(t_{tot}))$ is a Mittag-Leffler type attenuation,

$$\Upsilon(t_{tot}) = \frac{1}{\pi}\sqrt{\frac{\langle\phi_{Diff}\rangle^2}{2}}, \quad (68b)$$

and

$$\langle\phi_{Diff}\rangle^2 = [\langle\phi'_D(t_{tot})^2\rangle + \langle(\phi_D(t_{tot}))^2\rangle]. \quad (68c)$$

### 3. Results and discussions

The phase shift and signal attenuation for spin diffusion under a quadratic gradient field are derived. The recursive evaluation method is illustrated in detail, showing how to obtain various parameters such as the phase diffusion coefficient, phase variances, and phase shift from three different types of phase evolutions: phase diffusion, float phase, and phase shift.

From the recursive method, the float under $n$ order gradient field can be calculated based on the phase evolution of $n-2$ order gradient field. The float phase of the quadric field is thus calculated based on the phase diffusion results of the parabolic field, which corresponds to three components: $\phi'_D$, $\phi'_{float}$, and $\phi'_{shift}$ of the float phase of the quadric field. It is worth noting that in the recursive calculation, the virtual wavenumbers $K'_2(t_{tot})$ are not zero, which are $6D\gamma g_4\delta^2$ and $-12D\gamma g_4\Delta\delta$ for one gradient pulse case, and two gradient pulses, respectively; instead of $K'_2(t)$, $K'_2(t_{tot}) - K'_2(t)$ should be used in the evaluation of the float phase.

The phase float term can also be derived from asymmetric diffusion. Asymmetric diffusion occurs when the jump probability or jump length differs in various directions of the random walk. In addition to the jump probability or jump length, the difference in jump times influences the drift velocity and diffusion coefficient. In the phase diffusion equation (Eq. 2), the jump lengths vary depending on direction. When the jump direction of $\Delta z_{j+1}$ has the same or opposite signs as the current position $z(t_j)$, their jump lengths are slightly different; the same sign jump results in a slightly larger jump length than the opposite sign jump, leading to a net phase drift velocity according to Eq. (8), which provides a drift velocity formula for asymmetric diffusion.

A general average phase expression can be obtained for an $n$-order nonlinear gradient field from a real-space integral based on Eq. (62), which results in Eq. (63), $12\gamma g_4 \delta D^2 \Delta^2 + 12\gamma g_4 \delta D^2 \Delta \delta + 12\gamma g_4 \delta z_0^2 D\Delta$, when $n = 4$. Eq. (63) is equal to the sum of $\emptyset'_{float}(t_{tot}), \emptyset'_{shift,Z_0}(t_{tot})$, and $\emptyset_{shift,Z_0}(t_{tot})$ from Eqs. (34), (39), and (40), results of the phase diffusion method.

It is worth noting that when the phase distribution is asymmetric, $\langle \phi \rangle$ could be different from the real phase of the total magnetization. For even $n$, the phase distribution should be asymmetric, and the real phase of the observed signal is related to the angle determined by $arctan \frac{\langle sin(\phi) \rangle}{\langle cos(\phi) \rangle}$ of the observed magnetization vector. One of the merits of the phase diffusion method is that it provides a relatively clear physical picture of these three types of phase evolutions, which is not clear in the average phase obtained based on Eq. (62).

The float phase components $\emptyset'_{float}(t_{tot}), \emptyset'_{shift,Z_0}(t_{tot})$ described by Eqs. (34) and (39) are non-zero, which may not be desirable in real experiments. Designing the gradient pulse sequence by doubling the sequence with opposite gradient pulses could eliminate $\emptyset'_{float}(t_{tot})$ and $\emptyset'_{shift,Z_0}(t_{tot})$.

The comparison between theoretical calculation and simulation for spin self-diffusion in a one-pulse case is shown in Fig. 2. The random walk simulation is carried out with 20k random walks, and each random walk has a few thousand jumps [23,28]. In Fig. 2, the simulation values are somewhat larger than those of the theoretical predictions. One of the theoretical curves is drawn based on the phase variances from the diffusion parts, $\langle \emptyset_D^2 \rangle$ and $\langle \emptyset'_D{}^2 \rangle$, while another one also includes $\langle \emptyset'_{float}{}^2 \rangle$. The value difference between theoretical predications and simulation could arise from the correlations between these phase components. Additionally, phase diffusion consists of different diffusion components with varying jump lengths, where large jump lengths follow large jump lengths, while short jump lengths follow short jump lengths. Further research is required to understand these correlations.

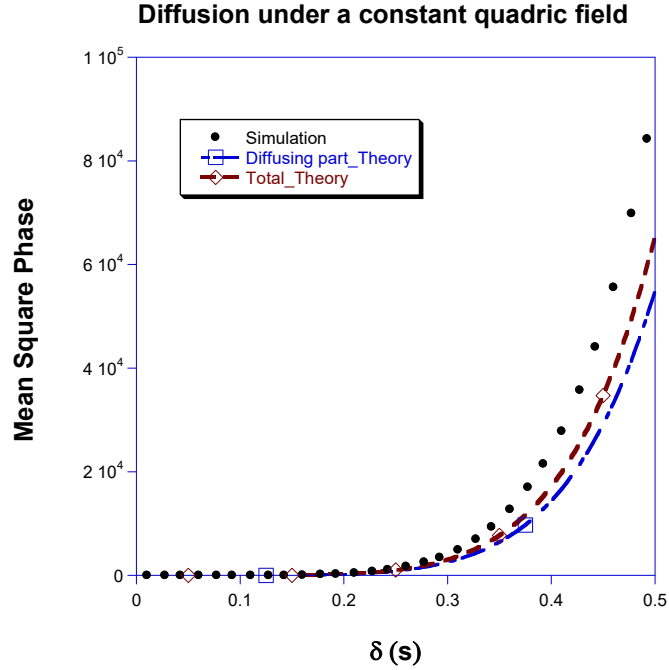

**Fig. 2** The phase variance for diffusion under a constant quadric field. The parameters $g_4$=1920 T/cm$^4$, $\gamma = 2.675 \times 10^8$ rad/s/T, and $D = 2 \times 10^{-9}$ m$^2$/s, are used. One of the theoretical curves is obtained based on the phase variances from the diffusion parts, $\langle \phi_D^2 \rangle$ and $\langle \phi'^2_D \rangle$, while another one includes $\langle \phi_D^2 \rangle$, $\langle \phi'^2_D \rangle$, and $\langle \phi'^2_{float} \rangle$.

A different type of attenuation function $\frac{1}{\sqrt[2n]{1+\varepsilon\langle\phi^2_{Diff}\rangle}}$ is proposed for signal attenuation in the nonlinear gradient field. For spins initiating diffusion from the origin of the nonlinear gradient field, the signal attenuation differs from the Gaussian diffusion type attenuation in a linear field, which exhibits exponential attenuation. $\frac{1}{\sqrt[2n]{1+\varepsilon\langle\phi^2_{Diff}\rangle}}$ is generalized from the SGP approximation, Eq. (60), for spin self-diffusion from the origin of a parabolic gradient field. This generalization includes two aspects: a. from the SGP approximation to include the FGPW effect, b. from order $n = 2$ to $n \geq 3$. The extension from SGP approximation to the FGPW effect is reasonable as the phase distributions are similar, resulting from the same types of phase evolution equation, Eq. (2). However, for $n \geq 3$, because the integrals in Eq. (59) currently do not have closed forms, the generalization from $n = 2$ to $n \geq 3$ is not based on a strict derivation, which is a guess based on comparing with numerical data. Further research is needed to optimize the expression.

The predicted signal attenuation based on Eq. (68a) is compared with simulations as shown in Fig. 3. The signal attenuation is Gaussian attenuation for small attenuation, then changes to other types of attenuations: Lorentzian, MLF attenuations (MLF is evaluated based on Pade approximation [29]), or extended SGP approximation. There are good agreements between the generalized SGP predication and simulation for a range from small attenuation to around 50% attenuation, in all three gradient sequences. The parameter $\varepsilon$ used is 2 for one-pulse, while ½ for two-pulse. Noting that the $\varepsilon$ value used is ¼ in SGP in

Fig. 1. The change of $\varepsilon$ from 2 to ¼ seems reasonable considering the two-pulse case is a combination of two one-pulses and SGP, which should have an intermediate value. The parameters of expression could be optimized through additional research. It is worth mentioning that the signal attenuation is Gaussian for spins starting from positions far away from the origin, as $\langle [z(t)]^n \rangle \approx (Z_0)^n$ when $(Z_0)^2 \gg 2Dt$ . Further effort is needed to better describe the signal attenuation, particularly, the large attenuation.

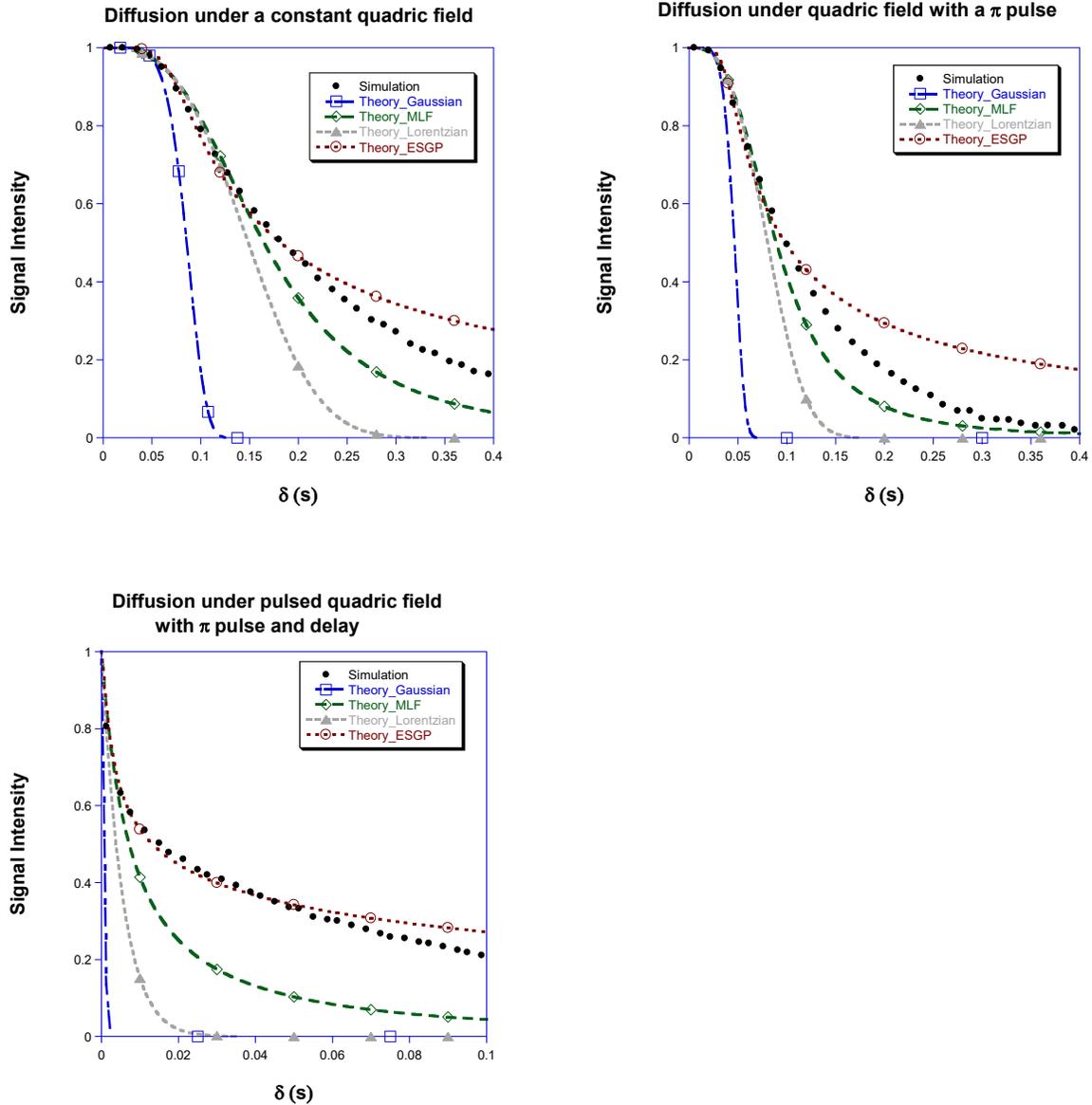

**Fig. 3** Comparison of theoretical NMR signal attenuations with random walk simulations. (a), (b), and (c) are for diffusion under different sequences. The parameters, $g_4$=1920 T/cm⁴, $\gamma = 2.675 \times 10^8$ rad/s/T, and $D = 2 \times 10^{-9}$ m²/s, are used. Additionally, $\Delta - \delta = 0.5$ s are used for (c). The signal attenuation obeys Gaussian attenuation for the short time, then changes to other types of attenuations: Lorentzian, MLF, or the extended SGP approximation.

The higher order of the gradient field provides unique benefits. As $\langle \phi^2 \rangle \propto D^n t^{n+2}$, the signal attenuation with higher orders is more sensitive to changes in the diffusion coefficient and diffusion delay, which could improve image contrast in MRI images. Additionally, the phase shift from the float phase could be employed to measure the diffusion coefficient directly [20].

The phase distribution and signal attenuation in a nonlinear gradient field are more complex than those in a linear gradient field. Theories such as the phase diffusion method [7,23] and the real-space methods, such as the phase shift average, SGP, and others in Refs. [11-14,21-22], approach this challenge from different perspectives. These results offer complementary insights that help advance nonlinear gradient research.

**References**

[1]  E.L. Hahn, Spin echoes, Phys. Rev. 80 (1950) 580-594.
[2]  H. C. Torrey, Bloch Equations with Diffusion Terms, Phys. Rev. 104(3) (1956) 563－565.
[3]  D. W. McCall, D. C. Douglass, E. W. Anderson, Ber Bunsenges Physik Chem. **1963**, *67*, 336-340.
[4]  E. O. Stejskal, J. E. Tanner, J. Chem. Phys. 42 *(*1965**)** 288-292; doi: 10.1063/1.1695690.
[5]  W.S Price, NMR Studies of Translational Motion: Principles and Applications; Cambridge University Press, 2009.
[6]  P. Callaghan, Translational Dynamics and Magnetic Resonance: Principles of Pulsed Gradient Spin Echo NMR; Oxford University Press, 2011.
[7]  G. Lin, An effective phase shift diffusion equation method for analysis of PFG normal and fractional diffusions, J. Magn. Reson. 259 (2015) 232–240.
[8]  G. Lin, Phase diffusion methods for NMR spin system evolutions: Diffusion, relaxation, and exchange, Annual Reports on NMR Spectroscopy, 114 (2025) 33-181.
[9]  D.W. McRobbie, E.A. Moore, M.J. Graves, M.R. Prince, MRI From Picture to Proton, 2nd ed., Cambridge University Press, Cambridge, England, 2007.
[10] K.F. Morris, C.S. Johnson Jr., Diffusion-ordered two-dimensional nuclear magnetic resonance spectroscopy, J. Am. Chem. Soc. 114 (8) (1992) 3139–3141.
[11] P. Bendel, Spin-echo attenuation by diffusion in nonuniform field gradients, J. Magn. Reson. 86 (1990) 509–515.
[12] P. Doussal, P. N. Sen, Decay of nuclear magnetization by diffusion in a parabolic magnetic field: an exactly solvable model, Phys. Rev. B 46 (6) (1992) 3465–3486.
[13] M.D. Huerlimann, Effective gradients in porous media due to susceptibility differences, J. Magn. Reson. 131 (1998) 232–240.
[14] O. Posnansky, R. Huang, N.-J. Shah, J. Magn. Reson 173 (2005) 1.
[15] W. C. Kittler, P. Galvosas, M. W. Hunter, Parallel acquisition of q-space using second order magnetic fields for single-shot diffusion measurements, J. Magn. Reson. 244 (2014) 46-52.
[16] R. Bammer, M. Markl, A. Barnett, B. Acar, M.T. Alley, N.J. Pelc, G.H. Glover, M.E. Moseley, Analysis and generalized correction of the effect of spatial gradient field distortions in diffusion-weighted imaging, MRM 50 (2003) 560–569.
[17] D. LeBihan (Ed.), Magnetic Resonance Imaging of Diffusion and Perfusion: Applications to Functional Imaging, Lippincott–Raven Press, New York, 1995.
[18] L.J. Zielinski, P.N. Sen, Relaxation of nuclear magnetization in a nonuniform magnetic field gradient and in a restricted geometry, J. Magn. Reson. 147 (2000) 95–103.

[19] K. Tušar, I. Serša, Use of nonlinear pulsed magnetic fields for spatial encoding in magnetic resonance imaging, Sci. Rep. 14, 7521 (2024). https://doi.org/10.1038/s41598-024-58229-x.

[20] P. Wochner, T. Schneider, J. Stockmann, J. Lee, R. Sinkus, Diffusion phase-imaging in anisotropic media using non-linear gradients for diffusion encoding, PLoS One. 18(3) (2023) e0281332. doi: 10.1371/journal.pone.0281332. PMID: 36996066; PMCID: PMC10062566.

[21] G. Lin, Z. Chen, J. Zhong, D. Lin, X. Liao, A novel propagator approach for NMR signal attenuation due to anisotropic diffusion under various magnetic field gradients, Chem. Phys. Lett. 335 (2001) 249–256.

[22] J.C. Tarczon, W.P. Halperin, Interpretation of NMR diffusion measurements in uniform- and nonuniform-field profiles, Phys. Rev. B 32 (5) (1985) 2798–2806.

[23] G. Lin, Effective phase diffusion for spin phase evolution under random nonlinear magnetic field, Phys. Rev. E 110(3) (2024) 034119.

[24] Chmurny GN, Hoult DI. The ancient and honourable art of shimming, Concepts Magn.Reson. 2(3) (1990) 131–149. https://doi.org/10.1002/cmr.1820020303.

[25] C. Améndola, J.-C. Faugère, B. Sturmfels, Moment Varieties of Gaussian Mixtures, J. Algebr. Stat, 7(1) (2016) 14-28.

[26] G. Lin, S. Zheng, Diffusion coefficient expression for asymmetric discrete random walk with unequal jump times, lengths, and probabilities, Phys. A 638 (2024) 129620.

[27] M. Holmes, Asymmetric random walks and drift-diffusion, EPL 102 (2013) 30005.

[28] G. Lin, J. Zhang, H. Cao, A. A. Jones, A lattice model for the simulation of diffusion in heterogeneous polymer systems. Simulation of apparent diffusion constants as determined by pulse-field-gradient nuclear magnetic resonance, J. Phys. Chem. B. 107 (2003) 6179–6186.

[29] C. Zeng, Y. Chen, Global Pade approximations of the generalized Mittag-Leffler function and its , Fract. Calc. Appl. Anal. 18 (2015) 1492–1506.